\documentclass[epj]{svjour}

\usepackage{graphicx}
\usepackage{amsmath,amssymb,amsfonts,mathrsfs}
\usepackage{epic,eepic}
\usepackage{graphics}
\usepackage{epsfig}
\renewcommand{\vec}[1]{\mathbf{#1}}

\usepackage{subfigure}

\def\id{{1 \kern-.30em 1}}

\usepackage{color}
\usepackage{graphicx}

\authorrunning{M. Sega et al.}
\begin{document}

\title{The importance of chemical potential in the determination of water slip in nanochannels}

\author{M. Sega\inst{1}, M. Sbragaglia\inst{2}, L. Biferale\inst{2}, S. Succi\inst{3}}
\institute{ Institute of Computational Physics, University of Vienna, Sensengasse 8/9, 1090 Vienna, Austria \and Department of Physics and INFN, University of ``Tor Vergata'', Via della Ricerca Scientifica 1, 00133 Rome, Italy \and Istituto per le Applicazioni del Calcolo CNR, Via dei Taurini 19, 00185 Rome, Italy }

\abstract{
We investigate the slip properties of water confined in graphite-like nano-channels by non-equilibrium molecular dynamics simulations, with the aim of identifying and analyze separately the influence of different physical quantities on the slip length. In a system under confinement but connected to a reservoir of fluid, the chemical potential is the natural control parameter: we show that two nanochannels characterized by the same macroscopic contact angle  -- but a different microscopic surface potential -- do not exhibit the same slip length unless the chemical potential of water in the two channels is matched. Some methodological issues related to the preparation of samples for the comparative analysis in confined geometries are also discussed. 
}

\maketitle
\section{Introduction}
\begin{figure}[t]
\begin{center}
\includegraphics[width=0.8\columnwidth]{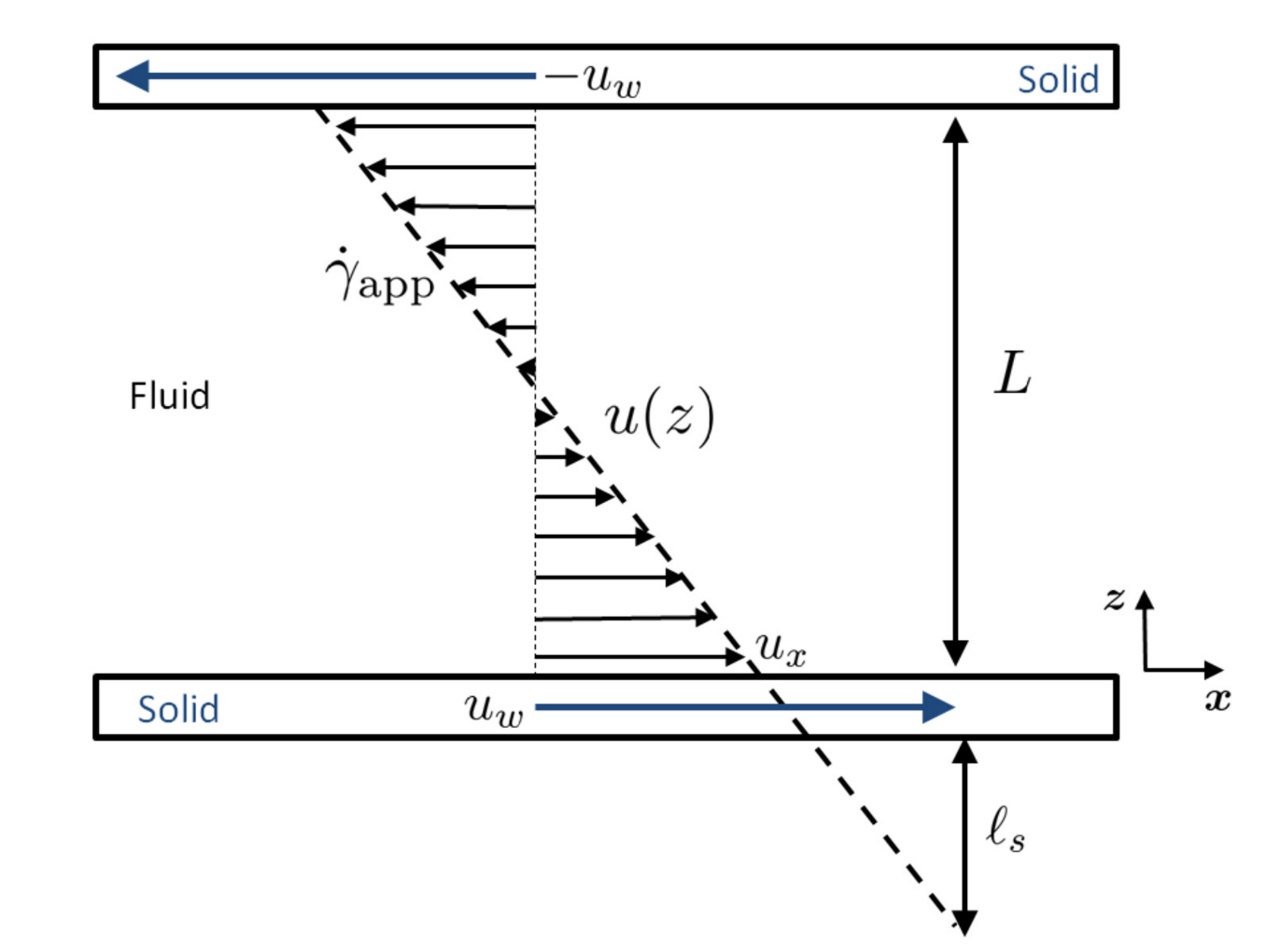}
\end{center}
\caption{A sketch for the steady Couette flow geometry with two solid walls moving with opposite velocities $\pm u_w$. A non zero slip length (see equation (\ref{eq:slip})) can be defined at the boundaries when a velocity difference develops between the solid ($u_w$) and the adjacent liquid layer ($u_x$). Consistently, an apparent shear rate $\dot{\gamma}_{\mathrm{app}} \neq 2u_w/L$ develops.\label{fig0}}
\end{figure}

The possibility of slip at the liquid-solid interfaces has been debated for over two centuries \cite{lamb}. The recent blossoming of research in micro- and nanofluidics has prompted renewed interest in the possibility of slip at the interface between a liquid and a solid \cite{stone04a,squires05a,sbragaglia05a,benzi06a,benzi06b}, since the classical no-slip boundary condition of macroscropic hydrodynamic theory does not necessarily hold at those scales \cite{tabeling09a,lauga07a}. Instead, the motion of a viscous fluid at a solid interface is well described by a partial-slip (Navier) boundary condition that relates the velocity difference between the solid ($u_w$) and the adjacent liquid ($u_x$) to its normal (say along $z$) gradient (see figure \ref{fig0})
\begin{equation}
u_x-u_w=\ell_s \left. \frac{\partial u_x}{\partial z} \right|_{z=0}. 
\end{equation}
The physical rationale behind the slip phenomenon is a balance between the frictional forces and the viscous forces. This is apparent in the very same definition of the slip length $\ell_s$ in a channel of width $L$ (see figure \ref{fig0})
\begin{equation}\label{eq:slip}
\ell_s = \eta/\lambda - L/2
\end{equation} 
where the viscosity $\eta=F/\dot{\gamma}_{\mathrm{app}}$ and the friction coefficient $\lambda=F/u_w$ \cite{Bocquet10} are determined by the force per unit surface $F$ between the wall and the fluid and the apparent shear rate $\dot{\gamma}_{\mathrm{app}}$. 
The role of interfacial slip becomes relevant in confined micro-
and nanofluidic geometries, when the surface to volume ratio increases
substatially \cite{tabeling09a,lauga07a}, so that the slip length
may become comparable with the system size, $\ell_s \approx L$: an
accurate understanding of nanoscale friction phenomena at the
fluid-solid interfaces is indeed paramount to the design of micro-
and nanofluidic devices aimed at optimizing mass transport against
an overwhelming dissipation barrier \cite{eijkel05,schoch08a,Bocquet10}.
In the recent years, progress has been made in providing a coherent
description of slip motion close to solid boundaries, both from the
experimental \cite{tabeling09a,lauga07a} and theoretical \cite{bocquet07}
side, and general consensus on the occurrence of slip on partially
wetting substrates has emerged. Nevertheless, the physical mechanisms
leading to fluid slip are still not completely understood
\cite{huang08b,ho11b,martini08a,priezjev06a}. In a recent paper,
Huang and coworkers \cite{huang08b}, using molecular dynamics
simulations of an atomistic water model, studied the interfacial
hydrodynamic slippage at various hydrophobic surfaces. The measured
slip lengths range from nanometers to tens of nanometers and were
shown to almost collapse onto a single curve as a function of the
static contact angle $\theta$ characterizing the surface wettability
\cite{degennes03a}, thereby suggesting a quasi-universal relationship
$\ell_s\propto(1+\cos \theta )^{-2}$ \cite{huang08b}, with $\theta$
the equilibrium contact angle formed by liquid droplets at the solid
surface.  Rationalizing this picture in terms of the statistical
properties of atomistic motion close to solid boundaries is a
challenging task
\cite{bocquet94a,martini08a,petravic07a,hansen11a,chinappi12a},
especially when different molecular mechanisms of slip may coexist
\cite{martini08a,pahlavan11}: depending on the external driving
force (the applied shear), liquid atoms may hop from one equilibrium
site to another of the liquid-solid energy landscape, or even display
a collective motion of entire layers of atoms slipping together.
More recent numerical simulations \cite{ho11b} even suggested that
the dichotomy between hydrophobicity and large slip might be purely
coincidental and that hydrophilic surfaces can show features typically
associated with hydrophobicity. The description of the fluid and
surface at atomistic level is essential for these investigations,
as small changes in the surface properties can lead to surprisingly
different structural and dynamical properties of the fluid, as it
is seen for example for hydroxylated surfaces \cite{ho11a,qiao12a}.

Another issue is that of the shear-rate dependence of the slip
length: Thomson and Troian discovered, for example, that the slip
length can follow a universal curve as a function of the applied
shear rate, leading to a divergence.\cite{Thompson97} Priezjev,
however, showed  that an important role in this situation is played
by static surface roughness\cite{Priezjev07a}. As it turned out,
displacements of the atoms in the surface layer as small as those
generated by thermal fluctuations in graphite (about one tenth of
an Angstrom) are enough to remove the slip length divergence,
demonstrating how sensitive this quantity is on the microscopic
detail of the surface\cite{sega13a}.

Further analysis on the molecular mechanisms responsible for these
stimulating results is needed: the picture of a quasi-universal
relationship between wettability and slip length is surely appealing,
but it should be born in mind that as a dynamical quantity, the
slip length can not depend only on static properties like the contact
angle (that is, on the interfacial free energies). Indeed, a quantity
with the dimension of a diffusion coefficient appears in the
approximate expression for the slip length obtained from Green-Kubo
relations \cite{huang08b,Sendner09}.  More detailed calculations
suggest this quantity to be related to the collective solvent
diffusion tensor \cite{barrat99a}.  This clarifies the meaning of
the quasi-universal relation proposed in \cite{huang08b}, which
strictly speaking holds true only if the dynamical properties of
the solvent in interaction with the different substrates do not
change much depending on the substrate nature. The presence of
marked surface inhomogeneities, the possibility of making hydrogen
bonds, or the presence of dissociated charged groups at the
solid-liquid interface can of course drastically change this picture.

In this paper we address this issue by extending the systematic
approach that allowed us to tackle the slip length divergence
problem\cite{sega13a}. The approach consists in employing model
surfaces with different microscopic properties (i.e. a different
interaction potential between solid wall particles and water), but
the same macroscopic contact angle of the solvent.  In our previous
work, however, we used the same number of water molecules in each of the simulations
with different surfaces, hence, water was simulated at different chemical potentials. Here,
we investigate the effect of this difference in chemical potential
on the slip length, and show that two different surfaces, even if
they share the same macroscopic contact angle, can indeed lead to
a considerably different slip lengths. This difference, however,
vanishes when the chemical potential of water in the two channels
is matched. This has some important implications from the practical
and from the theoretical point of view, which are here discussed.

\section{Methods and Systems}\label{sec:methods}

\begin{figure}[t]
\begin{center}
\includegraphics[trim=100 90 50 200,clip, width=1.0\columnwidth]{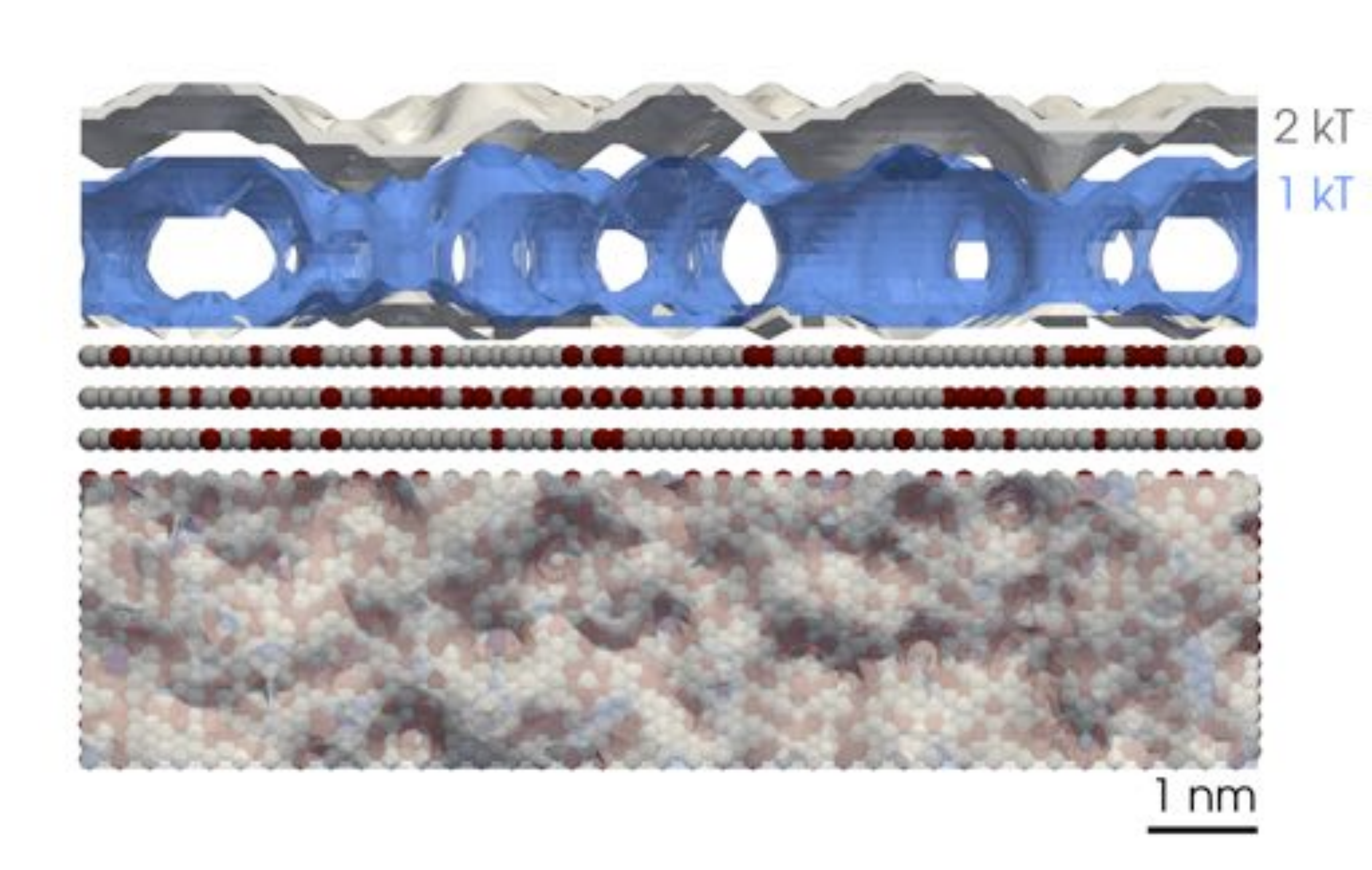}
\end{center}
\caption{
Top: side view of a detail of the three atomic layers composing the random quenched surface. Atoms with purely repulsive Lennard-Jones interaction are darker (red), while those with the enhanced attractive part of the potential are lighter (gray). On top of the atomic layers the equipotential surfaces at energy $E=kT$ and $E=2 kT$ are presented (not to scale in the direction perpendicular to the surface), to show the shape of the attractive basin next to the surface. Bottom: the top view of the detail of the random quenched surface. The atoms are depicted as in the top panel, and a translucent representation of the equipotential surface $E=kT$ is overlaid, in order to show the potential energy inhomogeneity.
\label{fig:surfacepot}}
\end{figure}

Using an in-house modified version of the Gromacs suite\cite{hess08a}, we simulated a slab of water molecules, modelled with the SPC/E potential \cite{berendsen87a} and confined between two graphite-like walls, each consisting of three atomic layers. The gap between the water-exposed layers is 6.982 nm, and the box edge sizes are 13.035,  16.188 and  8.0 nm in the $x$, $y$ and $z$ directions, respectively.

The electrostatic interaction was calculated using the smooth particle mesh Ewald method \cite{essmann95}, together with Yeh-Berkowitz correction \cite{yeh99} to remove the contribution from periodic images along the wall normal. Time integration of the equations of motion was performed using the leapfrog algorithm\cite{allen87a} with an integration time step of 1 fs. Water molecules are kept rigid using the SETTLE algorithm \cite{miyamoto92a}.  Both in equilibrium and non-equilibrium simulations, a Nos\`e--Hoover thermostat, coupled to the degrees of freedom not subjected to external forces, was employed to keep temperature constant. The atoms in the substrate are kept fixed in the equilibrium MD simulations, and are moved at a fixed velocity in the NEMD simulations. No center of mass removal procedure was applied.

We investigated two different setups, with respect to the wall-liquid interactions. In a {\it first} setup (here identified as ``standard'') the interaction potential of wall atoms with oxygen atoms is given by a Lennard-Jones potential
$U(r)=C_{12}r^{-12}-C_6 r^{-6} $, with  $C_6=2.47512\times 10^{-3}$ kJ/nm$^6$ and $C_{12}= 2.836328\times 10^{-06}$ kJ/nm$^{12}$\cite{vangunsteren96} up to a distance of $0.9$ nm. Above that distance, the potential  was smoothly switched to zero at $1.2$ nm, using 4th order polynomials. The {\it second} setup consists of a ``random quenched'' functionalization of the previous one, realized by making 40\% of wall atoms purely repulsive ($C_6=0$), and increasing the interaction strength of the remaining ones by a factor $\alpha$, to be tuned in order to achieve the same macroscopic contact angle as in the standard case. Equipotential energy surfaces and a snapshot of the three layers are shown in Fig.\ref{fig:surfacepot} for the random quenched case.

The procedure we used to investigate the slip length for different surfaces removing the biasing dependence on both contact angle and chemical potential, can be summarized as follows: a) we first generated the depleted surface by removing the attractive term of the Lennard-Jones interaction for a random selection of surface atoms; b) we calibrated the Lennard-Jones interaction strength $\alpha$ of the remaining atoms to obtain the same macroscopic contact angle of the standard surface case, using the generalized Young equation to extrapolate the contact angle to its macroscopic limit for a sessile droplet; c) we determined the number of water molecules that yields equal chemical potential for the systems with standard and depleted surfaces in a slit-pore configuration; d) we calculated the velocity profile induced by moving at constant speed and in opposite direction the two slabs of the slit pore, at a shear rate low enough to be in the constant slip length regime ; e) we eventually extracted the apparent slip length from a linear fit of the velocity profile, performed in the central part of the slit pore.

\section{Results and discussion}\label{sec:results}

\paragraph{The contact angle.} Given the known dependence of the slip length on the contact angle \cite{huang08b}, an equal wetting is a prerequisite to compare the slip length of water on two microscopically different surfaces and understand how much the slip length is affected by different types of surface inhomogeneities.
The actual value of the factor $\alpha$ has been chosen so that water wets the two surfaces with a comparable macroscopic contact angle.
To compute the macroscopic contact angle, we employed the procedure suggested by Werder and colleagues\cite{werder03}. For each  surface type, three droplets of different sizes (about $1.8\times10^3$,$14\times10^3$ and $34\times 10^3$ water molecules, respectively) on six monoatomic layers were simulated starting from a rectangular droplet shape, which is equilibrated for 500 ps (shape relaxation is observed to take place within few tens of ps).  Density profiles along the radial direction within slabs at different heights $z$, were sampled from the next 1 ns of simulation. A best fit to a sigmoid function is then employed to identify the location of the Gibbs dividing surface $R_{G}(z)$. A best fit to a circumference is eventually performed using the points of the locus $R_{G}(z)$, to extract the droplet radius $R$, base radius $R_B$ and contact angle $\theta$. The extrapolation of the contact angle to infinite $R_B$ is then performed via the linear fit o the generalized Young equation
\begin{equation}
\cos(\theta) = \cos(\theta_\infty) - \frac{\tau}{\gamma_{LV}}\frac{1}{R_B},
\end{equation}
where $\tau$ is the line tension, $\cos(\theta_\infty) = \gamma_{SV} - \gamma_{SL}$, and  $\gamma_{LV}$,  $\gamma_{SV}$  and $ \gamma_{SL}$ are the liquid-vapour, solid-vapour and solid-liquid surface tensions, respectively. The lateral size of the substrate layers is chosen to be at least a factor 1.5 larger than the droplet base radius.

\begin{figure}[t]
\begin{center}
\includegraphics[trim=65 0 55 0, clip, width=1.0\columnwidth]{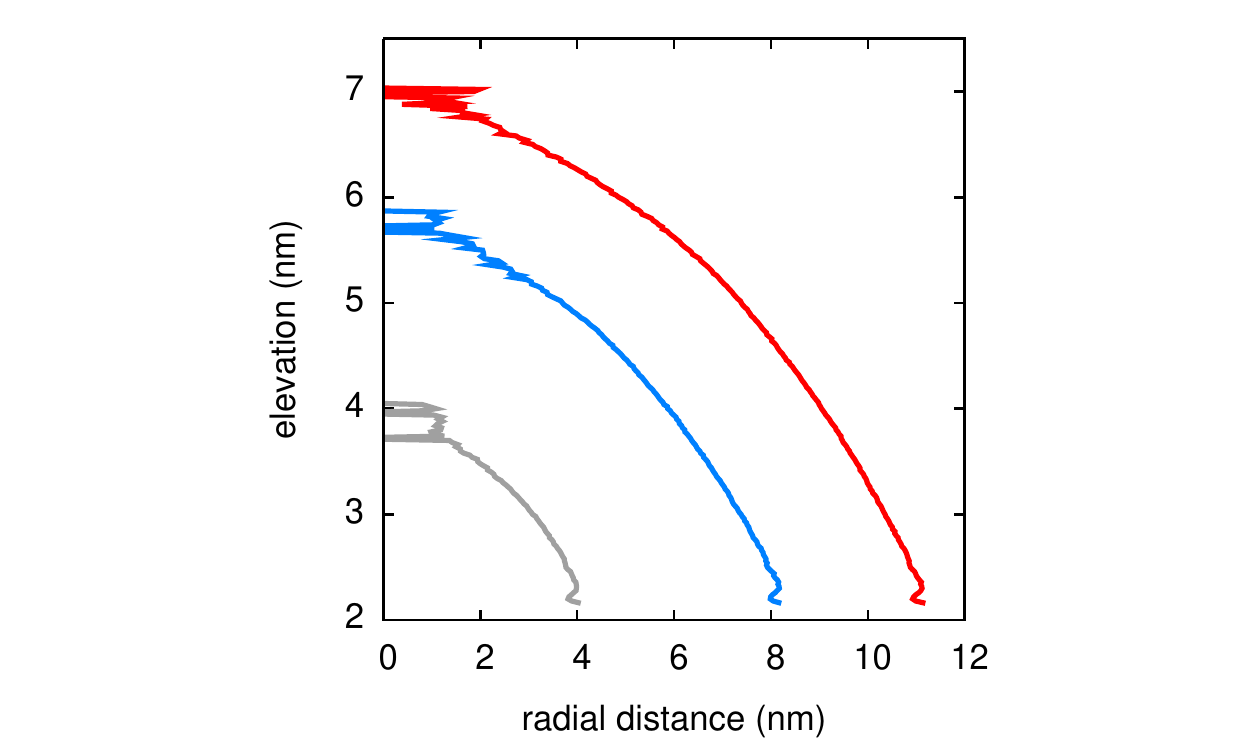}
\includegraphics[trim=0 500 0 400 ,clip,width=0.95\columnwidth]{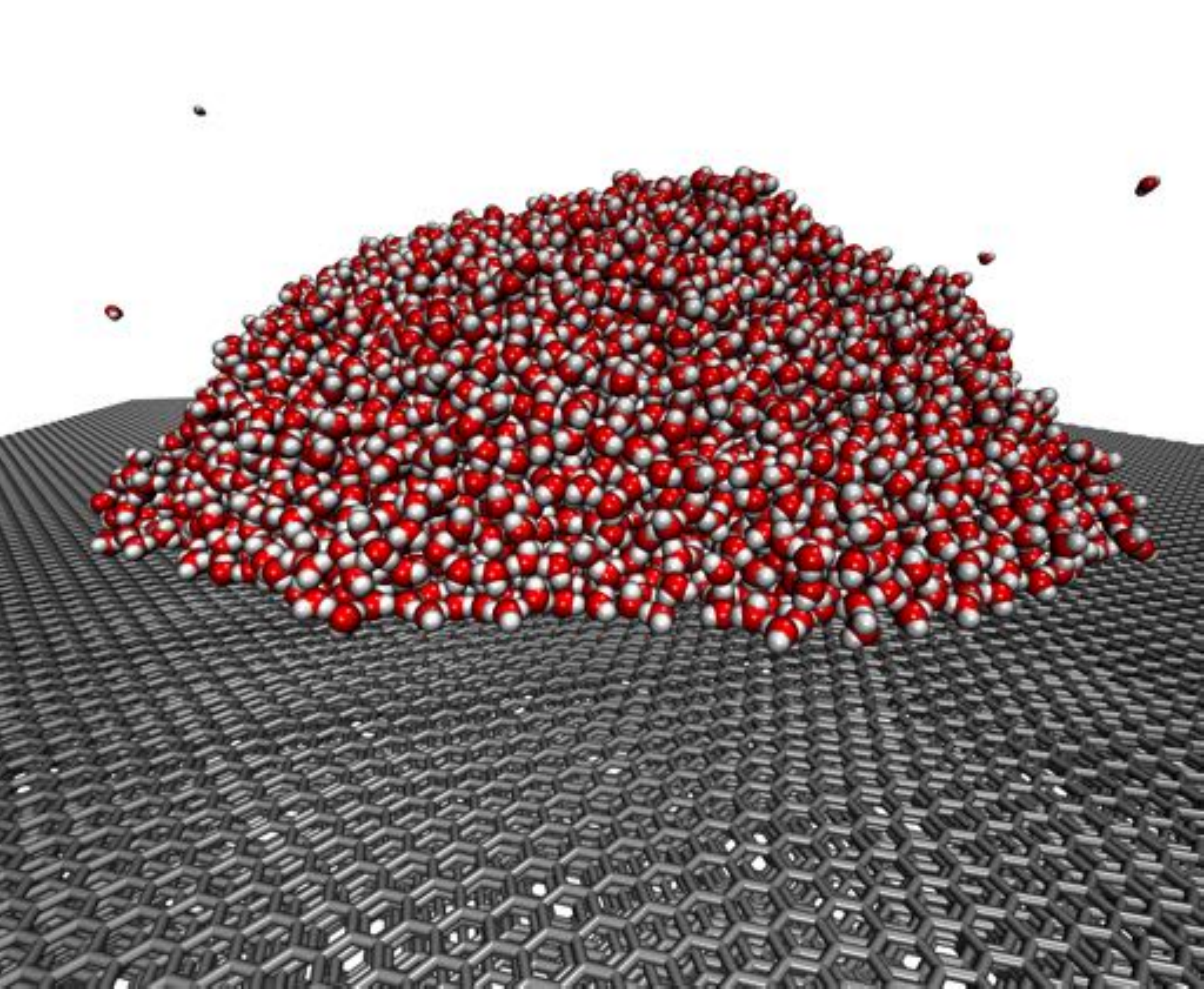}
\end{center}
\caption{Top: radial profiles of three different water droplets on the standard surface. The large fluctuations next to the top of the droplet are caused by the small volume element, and consequent low number of molecules associated to it. Bottom: simulation snapshot of one of the droplets, showing also some molecules in the vapour phase.
\label{fig:profiles}}
\end{figure}

\begin{figure}[t]
\begin{center}
\includegraphics[trim=25 0 8 0 ,clip,width=1.0\columnwidth]{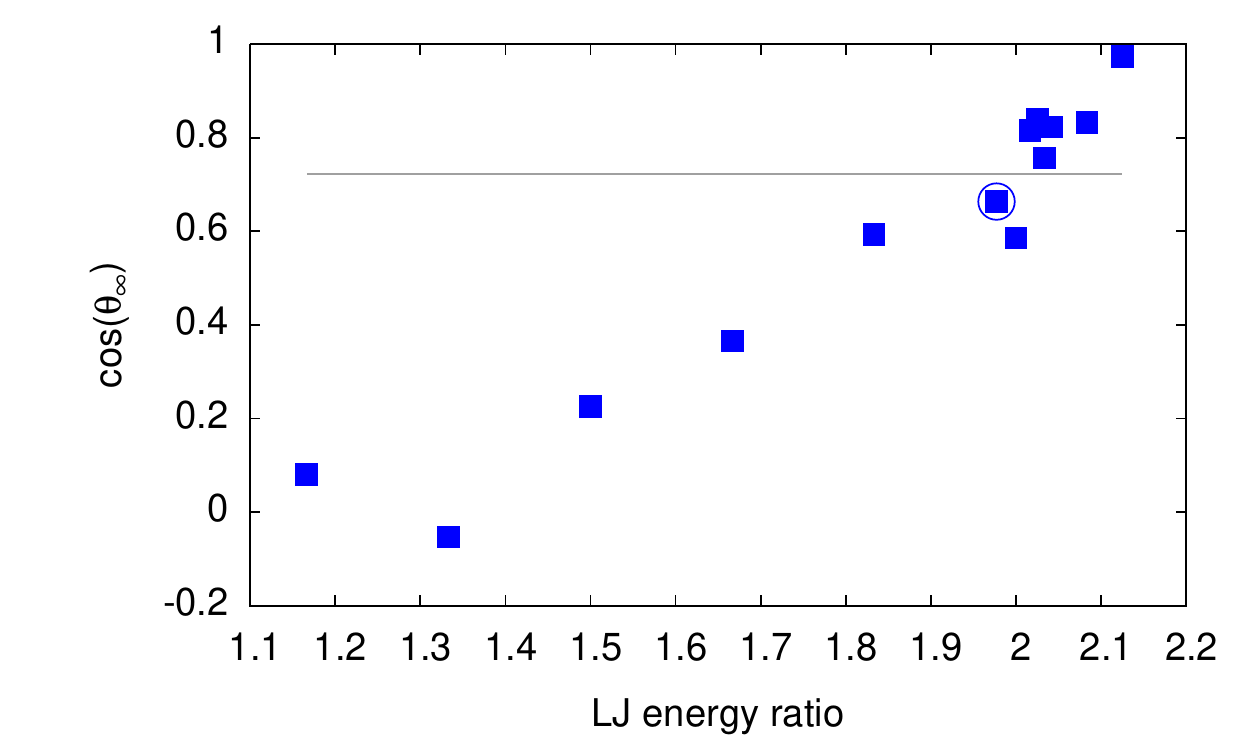}
\end{center}
\caption{Cosine of the macroscopic contact angle as a function of the ratio between the Lennard-Jones parameters $\epsilon=C_6^2/(4C_{12})$ of the atoms from set $S_2$ and that of the original force-field parameters. The value of the macroscopic contact angle for the standard surface is marked as a horizontal line. The circle on the point at a ratio of 1.977 marks the case chosen for the subsequent steps.
\label{fig:theta}}
\end{figure}

The calibration of the Lennard-Jones interactions turned
out to be a delicate procedure.  The cosine of the macroscopic contact angle, reported in Fig.\ref{fig:theta}, presents
a rather pronounced scattering, arising from the fitting of the radius $R_B$ and of the angle $\theta$, as well from the extrapolation to macrosopic droplet sizes,  which limited the accuracy in the determination of the macroscopic contact angle to about $\pm 5$ deg.
This limitation, however, turned out not to be overly restrictive for
our purposes, as the contact angle for the standard surface is about 40
deg, and the slip length should show only  a weak dependence on the contact
angle, for low values of the latter. In this case, the deviation
of the slip length due to a contact angle change of 5 deg would be
roughly 6\%.  As an outcome of this calibration procedure, we have
chosen the interaction of that 60\% of atoms that still retain the
attractive part of the potential to be a factor $\alpha=1.977$
stronger than in the case of the standard surface.

The cosine of the macroscopic contact angle, as it is seen in
Fig.\ref{fig:theta}, shows a linear dependence as a function of the
scaling factor applied to the Lennard-Jones interaction of the
attractive sites. A mean field approach is enough to explain this
behaviour, as it is known that the liquid-solid surface tension in
a fluid that interacts through the Lennard-Jones potential with the
surface depends linearly on the Lennard-Jones energy\cite{rowlinson}.
It is worth noticing that also the values of the line tension $\tau$,
although characterized by a high spread, can still be described
reasonably well by a linear dependence (reduced $\chi^2=0.8$) on
the interaction strength. The line tensions ranged from negative
values ($\tau\simeq$ -30 kJ/mol/nm for the smallest interaction
strength), to positive ones ($\tau\simeq$ 80 kJ/mol/nm).

\begin{figure}[t]
\begin{center}
\includegraphics[trim=22 20 10 5,clip,width=1.0\columnwidth]{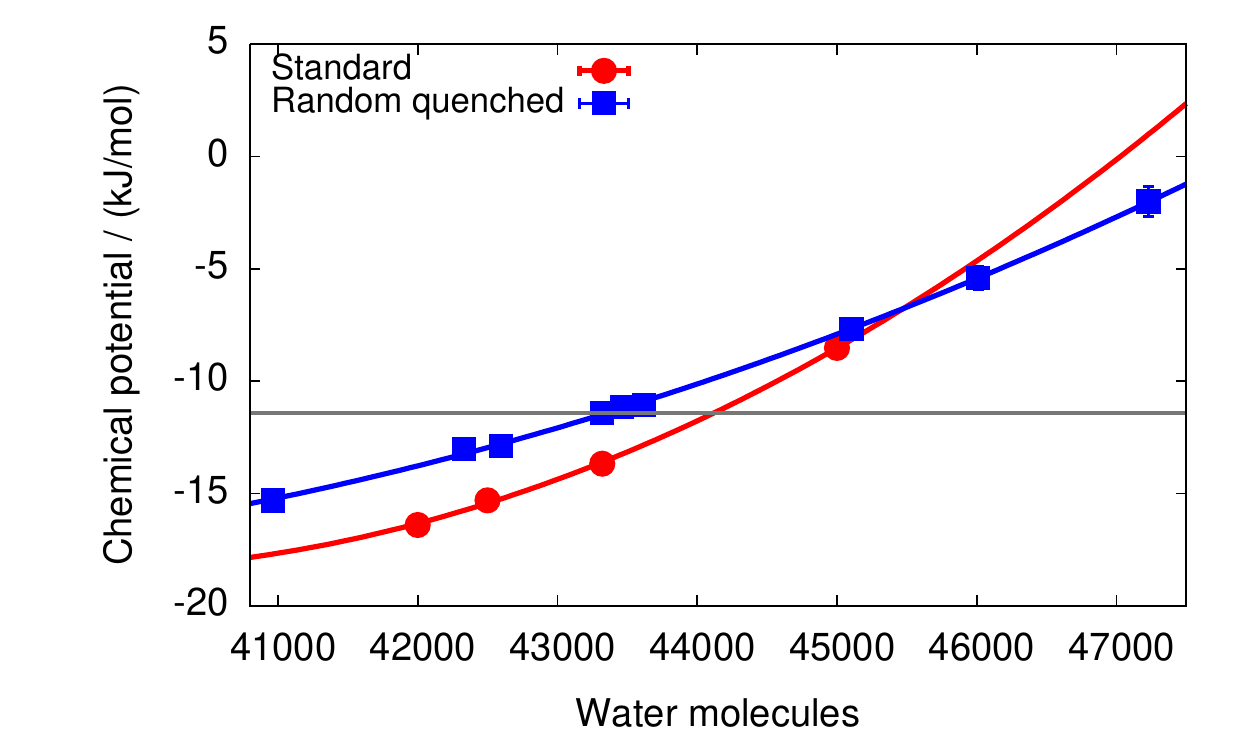}
\end{center}
\caption{Chemical potential of water confined between two slabs (circles, standard substrate; squares, depleted substrate). Solid lines are the result of a best fit to a quadratic function.\label{fig:mu}}
\end{figure}

\paragraph{The chemical potential.} Before being able to carry out the Couette flow simulation for the modified surface, the proper water content between the two slabs had to be determined.  The natural thermodynamic control parameters for a fluid confined between two flat interfaces are the accessible volume $V$, the temperature $T$, and the excess chemical potential $\mu_x$ of the fluid, since the latter is usually at thermal and chemical equilibrium with a reservoir. A series of MD simulations of a slit pore with different water content were hence performed, in order to identify the number of water molecules $N_w$ that leads to the same excess chemical potential of the standard surface case. The chemical potential was computed using the Widom insertion method \cite{widom63a} in its variant for inhomogeneous systems \cite{widom78a}. The chemical potential of a system with $N_w$ water molecules at one point $\vec{r}$ in space is given by
\begin{equation}
\mu_x(\vec{r}) = -\frac{1}{\beta} \ln\left[\frac{ \left\langle \exp(-\beta\Delta U(\vec{r}))\right\rangle  }{\rho(\vec{r})} \right],
\label{eq:widom}
\end{equation}
where $\Delta U$ is the change in potential energy due to the insertion of an additional molecule at position $\vec{r}$ with random orientation, and the canonical average was performed by sampling equilibrium configurations of the system with $N_w$ water molecules. For the Widom method, water molecules were inserted in those region where the density of water ensured proper statistics, namely, not too close to the surface. At chemical equilibrium, $\mu_x(\vec{r})$ has to be equal at every point in space, and this can be exploited to sample the chemical potential with high accuracy and relatively little computational effort.  In the slab system,  translational invariance along the $x$ and $y$ directions can be assumed --  not too close to the surface -- even if the surface is not homogeneous, so that the test molecule can be inserted at random positions on planes at different heights $z$. 
For the chemical potential calculation, the systems with different water content were simulated at equilibrium for 40 ns, saving configurations every 1 ps for the Widom test particle insertion analysis. For each frame stored, $10^4$ positions in a plane parallel to the substrate surface were randomly chosen as insertion sites for a test water molecule, and the energy difference was sampled according to the Widom scheme to compute the chemical potential. The obtained chemical potentials are reported in Fig.\ref{fig:mu}, together with the result of a fit to a quadratic function. The chemical potentials of the two systems, for a water content lower than $44\times10^3$ differ by about 2 kJ/mol, but become comparable in the proximity of  $N_w\simeq45\times10^3$. Since in this work we did not use any third reference system (open reservoir), the water content of one system can be chosen arbitrarily, and that of the other system can be changed to match the chemical potential.  We hence decided to use $N_w=43413$ for the standard surface, and, making use of the quadratic interpolation, we identified the value that yields the same chemical potential ($\mu_x=-11.4\pm0.12$  kJ/mol) for the depleted surface, namely, $N_w=44133$. The two corresponding densities will be denoted as $\rho_1$ and $\rho_2$, respectively ($\rho_2 > \rho_1$).

\begin{figure}[t]
\begin{center}
\includegraphics[trim=25 10 30 5,clip,width=1.0\columnwidth]{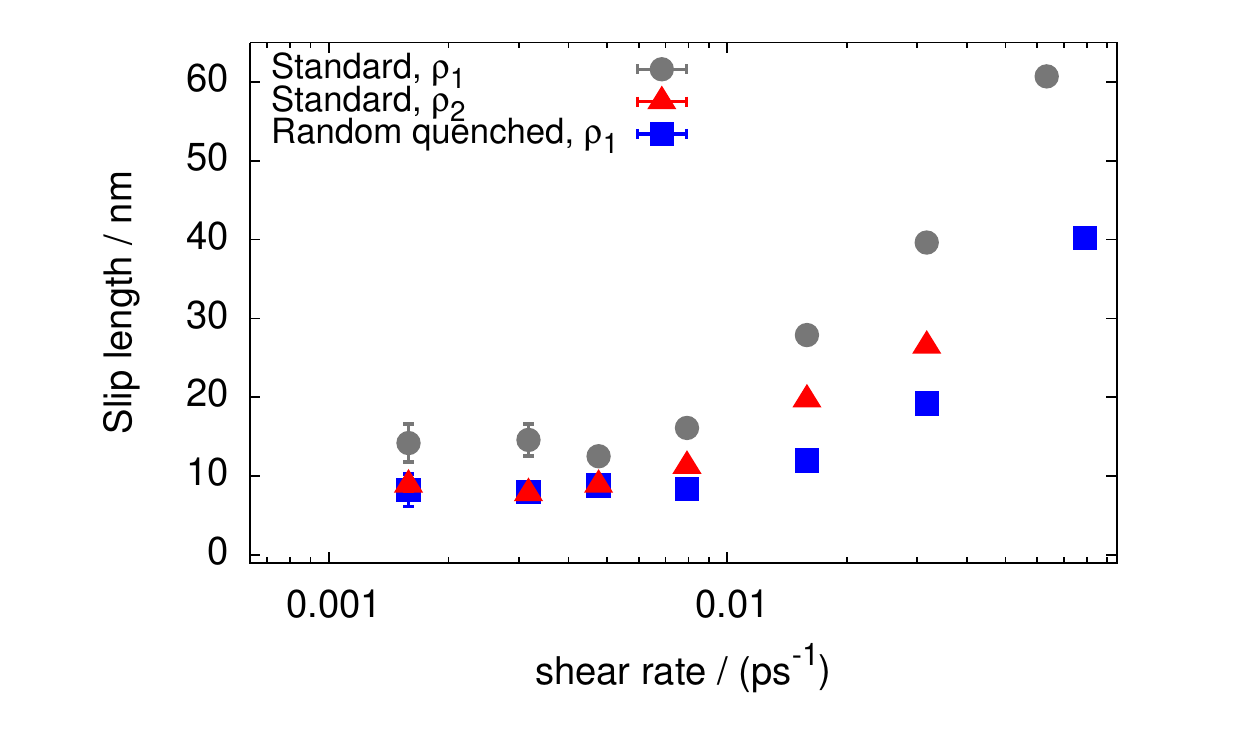}
\end{center}
\caption{Slip length of water, as a function of the imposed shear rate, for three different systems. Circles: standard surface with water density $\rho_1$. Squares: random quenched surface with water density $\rho_1$. Triangles: standard surface with water density $\rho_2$ (and with chemical potential matching that of squares). \label{fig:branches}}
\end{figure}

\begin{figure}[t]
\begin{center}
\includegraphics[trim=25 05 10 5,clip,width=1.0\columnwidth]{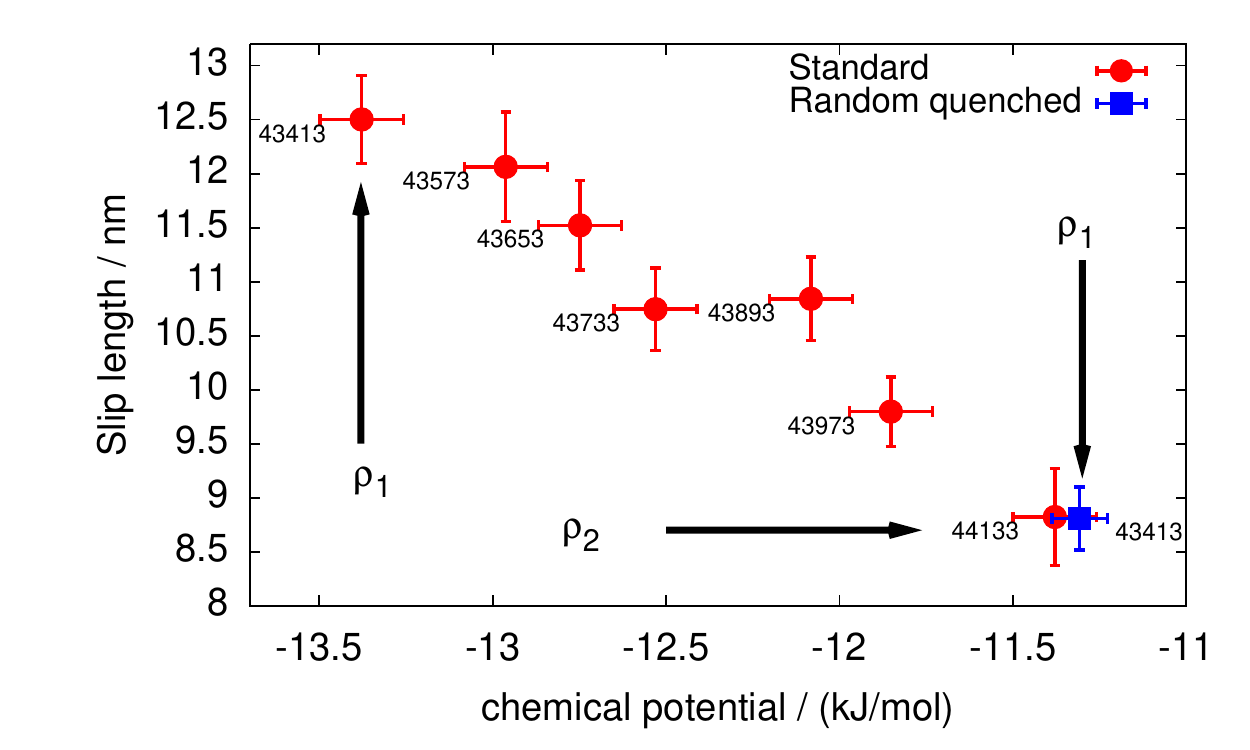}
\end{center}
\caption{
{ 
Slip length of water in the pore with standard surface, as a function of
the chemical potential (from the best fit), obtained for an imposed shear rate of 0.0047 ps$^{-1}$.  The water content $N_w$ of the channel is reported on the plot next to each point. The lowest and highest water content correspond to the densities $\rho_1$ and $\rho_2$, respectively. At the density $\rho_2$ the chemical potential of the standard surface system matches that of the random quenched surface system with water density $\rho_1$ (square)}
\label{fig:slip-mu}}
\end{figure}

\paragraph{The slip length.} Slip lengths are eventually calculated by extrapolating the velocity
profile of a Couette flow induced by imposing constant velocity on
two parallel, identical slabs made of the atoms of the standard or
depleted surface.  Configurations are saved every 1 ps to analyze
the velocity profile ($x$ component) along the $z$ direction and
to extract the apparent slip length, using a linear fit of the
resulting Couette flow. In Fig.\ref{fig:branches} we show the
slip length obtained for three different systems at various shear
rates. The three systems are: (i) the random quenched surface channel
with water density $\rho_1$ (squares); (ii) the standard,
unmodified surface with either (ii) $\rho_1$ (circles) or (iii)  $\rho_2$ (triangles) 
For all three cases, the slip length starts increasing noticeably
at shear rates higher than $\simeq 0.01$ ps$^{-1}$, but is
shear-independent below shear rates of about $0.005$ps$^{-1}$. In
this plateau region the systems with same water content but different
surface potential show a noticeably different slip length. On the
contrary, matching the chemical potential drives the system to the
same slip length (always in the plateau region). This result, we
would like to stress, is valid under the condition that the macroscopic
contact angle for water droplets is the same for both surfaces.

By calculating the slip length for the systems with the standard
surface at different water content it is possible to investigate
quantitatively the dependence on the slip length on the excess
chemical potential, as shown in Fig.~\ref{fig:slip-mu}. Even though
the statistical uncertainty is rather large, one can observe a
definite trend. The magnitude of the slip length change is particularly
important, as it passes from 9 to 12.5  nm (an increase of roughly
30\%, compared to a decrease of about 15\% in chemical potential).  
To provide some practical terms of comparison, one can consider
that the temperature coefficient for the chemical potential of water
at standard conditions is $\partial\mu/\partial T= -69.9$
J/mol/K\cite{job2006chemical}, hence, an equivalent change of 2
kJ/mol in the chemical potential of water could be realized, for
example, by raising its temperature by about 30 K.

It is an important question, wheter a change from 9 to 12.5~nm is
a significant one. In a Poiseuille (cylindrical) flow setup, the relative rate of work
$W_1/W_2$ necessary to achieve a target flux $Q$ in two systems characterized by
the slip lengths $\ell_1$ and $\ell_2$, respectively, is \begin{equation}
\frac{W_1}{W_2} = \left( \frac{  1+ 4 \ell_2 / R    } {  1+ 4
\ell_1 / R } \right)^2, \label{eq:ratio}\end{equation} where $R$ is the tube radius. This result can be obtained considering that the work rate (per unit length) in presence of a slip velocity $u(R)$ is
\begin{equation}
W =  \frac{\partial p}{\partial z} Q \left\{  1 - \frac{\pi R^2 u(R)}{Q} \right\},
\end{equation}
where $\partial p / \partial z $ is the pressure gradient. The first
term on the right hand side comes from the fluid deformation, while
the second one from the friction at the boundary\cite{landauFluids},
and can be cast in this form noticing that the friction force at
the surface has to balance the force generated by the pressure
gradient, $2\pi R \sigma = \pi R^2 \partial p / \partial z $, where $\sigma$ is the stress tensor.
For a Poiseuille flow, the slip velocity is 
\begin{equation}
u(R)=\frac{\partial p }{\partial z}\frac{R}{2\eta} \ell_s
\end{equation}
and the relation between pressure difference and flux is 
\begin{equation}
Q = \frac{\partial p}{\partial z}\frac{\pi R^4}{8\eta} \left(1+4\frac{\ell_s}{R}\right),
\end{equation}
from which, after a bit of algebra, Eq.~\ref{eq:ratio} can be derived.
With the present values of slip length, $\ell_1=12.5$, $\ell_2=9$, and using $R=L$, the outcome is a significant 
reduction, $W_1\simeq 0.54 W_2$, of the work rate necessary to sustain the flow.

\bibliographystyle{epj}
\bibliography{droemu}

\end{document}